\documentclass[runningheads,a4paper]{llncs}
\usepackage{multirow}
\usepackage[numbers]{natbib}
\usepackage{xr}
\usepackage{nameref}
\RequirePackage[textsize=scriptsize]{todonotes}
\usepackage{flushend}
\usepackage[textsize=scriptsize]{todonotes}
\usepackage{mdframed}
\usepackage{multirow,slashbox,pbox,booktabs}
\usepackage[export]{adjustbox}
\usepackage{pgfplots}
\usepackage{algorithmicx}
\usepackage[normalem]{ulem}
\usepackage[noend]{algpseudocode}
\usepackage{enumitem}
\usepackage{amssymb}
\usepackage{graphicx}
\usepackage{xcolor, colortbl}

\usepackage[hyphens]{url}
\urldef{\mailsa}\path|{alfred.hofmann, ursula.barth, ingrid.haas, frank.holzwarth,|
\urldef{\mailsb}\path|anna.kramer, leonie.kunz, christine.reiss, nicole.sator,|
\urldef{\mailsc}\path|erika.siebert-cole, peter.strasser, lncs}@springer.com|    
\newcommand{\keywords}[1]{\par\addvspace\baselineskip
\noindent\keywordname\enspace\ignorespaces#1}
\def\etal{\textit{et al.}}

\begin{document}

\mainmatter  
\title{The Evolving Ecosystem of Predatory Journals: A Case Study in Indian Perspective}
\titlerunning{The Evolving Ecosystem of Predatory Journals}

\author{Naman Jain, Mayank Singh}
\authorrunning{Naman Jain et al}
\institute{Department of Computer Science and Engineering\\ Indian Institute of Technology Gandhinagar, GJ, India \\ \{naman.j, singh.mayank\}@iitgn.ac.in }
\maketitle

\begin{abstract}
Digital advancement in scholarly repositories has led to the emergence of a large number of open access predatory publishers that charge high article processing fees from authors but fail to provide necessary editorial and publishing services. Identifying and blacklisting such publishers has remained a research challenge due to the highly volatile scholarly publishing ecosystem. This paper presents a data-driven approach to study how potential predatory publishers are evolving and bypassing several regularity constraints. We empirically show the close resemblance of predatory publishers against reputed publishing groups. In addition to verifying standard constraints, we also propose distinctive signals gathered from network-centric properties to understand this evolving ecosystem better. To facilitate reproducible research, we shall make all the codes and the processed dataset available in the public domain.  

\end{abstract}

\section{Introduction}
Scholarly journals play an essential role in the growth of science and technology. They provide an information sharing platform for researchers to publish and access scientific literature. However, high monthly access costs, pay-per-view models, complicated and lengthy publication process restrict researchers to leverage such knowledge. Open access journals (OAJ) emerged as a solution to abolish high access charges, with provision for unrestricted access to the latest findings. OAJs operate by charging article processing fees from the authors with a promising commitment of rigorous peer-review to maintain the quality and academic standard. The evolution of OAJs has been under investigation for at least a decade. Many fascinating studies have shown evidence of malpractices in large-scale. Majority of the research is involved in proposing a set of criteria to identify such malpractices and blacklist those violating these criteria. We review a series of such criteria and show that malpractices have gradually evolved and are difficult to detect through standard small-scale manual studies. We conduct empirical experiments to present 'facts' that clearly demarcate reputed journals from dubious journals.
\paragraph{\textbf{Publication ethics and malpractices}}
Identification of malpractices in publishing is a challenging yet an important problem. Several small-scale manual studies have been performed to identify these publishers and blacklist them. Xia~\etal~\cite{xia2017overview} present evidence of malpractices in OAJ's fee-based model. They observed that majority of OAJs use \textit{'enticing websites'} to attract researchers with a promise of pretentious peer-review and faster publishing process. Recently, Gopalkrishnan~\etal~\cite{gopalakrishnan2016india} verified above claims by surveying 2000 authors that published in dubious journals. Due to a high disparity in the fee structure and other popular competitors, several OAJs adopted malpractices such as false editorial board, poor peer-review, and incorrect indexing information~\cite{beall2012predatory}. Some recent and more advanced malpractices include the formation of \textit{'citation cartels'}~\cite{fister2016toward} among journals to elevate the impact factor. 
Sorokowski~\etal~\cite{sorokowski2017predatory} demonstrated malpractices in recruiting editors by creating a fake profile of a researcher and applying for an editorial position at 360 journals. To their surprise, 48 bogus journals accepted the application without even verifying the information. Phil Davis~\cite{davis2009open} submitted articles --- generated by a popular scholarly tool $SCIgen$ --- in several lousy quality journals which claim to perform peer-review for nominal processing charges. SCIgen is a program that generates random Computer Science research papers, including graphs, figures, and citations. Following a similar methodology, Bohannon~\cite{bohannon2013who9s} submitted grammatically incorrect papers to  304 OAJs. Out of 304 journals, 157 accepted the paper, 98 rejected, and the rest of them consider it for review or failed to respond. Most of the journals accepted articles based on formatting, layout, and language highlighting the malpractices in the peer-review process.

\paragraph{\textbf{Predatory publishing and growth}}
Beall~\cite{beall2012predatory} coined the term \textit{'predatory journals'} for those journals that charge high article processing fees from authors but fail to provide necessary editorial and publishing services. He proposed a list of subjective criteria for predatory classification based on a manual study performed on a limited set of journals. In addition, Beall curated a list of questionable journals and publishers (popularly known as 'Beall's list') which includes hijacked journals, fake-metric companies, and predatory journals or publisher. 
Shen~\etal~\cite{shen2015predatory} conducted a manual study of 613 journals from Beall's list and showed high growth in popularity within a span of five years. The publication count increased from 50,000 articles (in 2010) to 420,000 articles (in 2014).

\paragraph{\textbf{Limitations in current research}}
Majority of the research in identification and curation of dubious journals revolves around Beall's list. However, multiple works have shown limitations in this approach. Sorokowski~\etal~\cite{sorokowski2017predatory} criticized Beall for relying on editorial appeal rather than analyzing the published content. Laine~\etal~\cite{laine2017identifying} claimed that Beall's criteria list is not ranked in a preference order resulting in a geographical bias towards reputed journals being categorized as predatory from developing countries. The list of journals used by Xia~\etal~\cite{xia2017overview} and Gopalkrishnan~\etal~\cite{gopalakrishnan2016india} were sampled from the geographically biased Beall\'s list. Sorokowski~\etal~\cite{sorokowski2017predatory} criticized Bohannon~\cite{bohannon2013who9s} for targeting only specific journals and for not comparing both reputed and predatory journals. Moreover, Beall's list is currently discontinued due to unknown reasons~\cite{straumsheim_2017,raniwala_2018}. In addition, majority of the Beall's criteria have been bypassed by the evolving predatory ecosystem which we have discussed in Section \ref{sec:evol} through empirical experiments.

\paragraph{\textbf{Landscape of Predatory Journals in India}}
David Moher~\cite{shamseer2017potential} claimed that 27\% of world's predatory journal publishers are located in India and ~35\% corresponding authors are Indians. A similar study by Demir~\cite{demir2018predatory} on 832 predatory journals yields 62\% journals being located in India along with authorship and editorship contribution of 10.4\% and 57\% respectively. In July 2018, leading Indian newspaper, The Indian Express, published an investigative story~\cite{yadav_2018} claiming that Hyderabad is the Indian hub of predatory journals, operating more than 300 companies and publishing close to 1500 journals. A similar study by Patwardhan~\etal~\cite{patwardhan2018critical} claimed that $\sim$88\% of the journals present in the University Grants Commission of India (UGC)'s \textit{'approved list of journals'} could be of low quality. Following widespread criticism, UGC has removed 4,305 journals that are either predatory or low quality.

\paragraph{\textbf{Our Contribution}}
In our current study, we present an empirical study of one of the most popular Indian publishing group \textbf{OMICS} (OPG) that is largely considered as 'predatory' by several scholarly organizations and investigative journalists including The Guardian~\cite{hern_2018}, Federal Trade Commission (USA)~\cite{ftc_site} and The Indian Express~\cite{yadav_2018}. We present several anecdotal evidences to show how OPG dubiously bypasses the majority of Beall's criteria and cautiously adapting itself to resemble highly reputed publishing groups. As an interesting example, a comparison study with BioMedical Central (BMC) reveals that OPG shares several characteristics with reputed publisher. In addition, BMC follows multiple predatory criteria   proposed by Beall's. Through our findings, we propose several empirical verification strategies to find statistical evidence in decision making. We claim that similar strategies can be applied to any scholarly publishers.

\section{Datasets}\label{sec:datasets}
Understanding the predatory nature of open-access scientific journals requires a rich bibliographic dataset. We, therefore, downloaded the entire OMICS publishing group (OPG) dataset~\cite{omics_site}. It spans multiple subjects from broad research domains such as medical sciences, economics, and engineering and indexes 782 journals. OPG operates in 10 different publishing classes such as iMedPub LTD, Pulsus Group, and Allied Academics~\cite{wiki_omics}. Table~\ref{tab:classes} presents journal and article distribution in OPG classes. Majority of the articles belong to OMICS class followed by RROIJ, Imedpub, and TSI. Table~\ref{tab:dataset} presents general statistics of OPG, OMICS class (the parent group) and well-known BioMed Central publishing group. BMC~\cite{bmc_site}, established in 2000, is a prominent United Kingdom-based publishing group. It is an open access group owned by Springer Nature. Currently, it publishes more than 300 scientific journals. We claim that similar study can be conducted for other reputed open access publishers such as PLOS, Nature, and  Science. 

\begin{table}[!tbh]
\begin{minipage}[b]{.5\textwidth}
  \centering
\resizebox{\hsize}{!}{
\begin{tabular}{|c||c|c|c|c|}
\hline
\bf OPG classes & \bf Journals & \bf Articles & \bf SJR\\
\hline
OMICS & 498 & 70,247 & 15\\ \hline
Imedpub LTD     & 144 & 13,050 &1\\\hline
SciTechnol      & 55  & 4,734   & 0  \\\hline
RROIJ        & 36  & 18,703  & 0\\\hline
Trade Science Inc (TSI)          & 22  & 12,823  & 1\\\hline
Allied Academics      & 9   & 145   & 0 \\\hline
Open Access Journals  & 5   & 2,310  & 1 \\\hline
Scholars Research Library & 4   & 399   & 0 \\\hline
Pulsus       & 3   & 91    & 3   \\\hline
Andrew John Publishing    & 2   & 12   &0\\
\hline\hline
\bf Total & \bf 782  & \bf 122,514  & \bf 21  \\\hline
\end{tabular}}
 \caption{Journal and article distribution in OPG classes (second and third column). Fourth column represents per class count of OPG journals in SJR. Only 21 ($\sim2.6\%$) journals in OPG are indexed in SJR}
 \label{tab:classes}
 \end{minipage}
  \qquad
  \begin{minipage}[b]{.45\textwidth}
  \centering
  \resizebox{\hsize}{!}{
  \begin{tabular}{lcc} 
   &\textbf{OPG}&\\\cline{1-2}
   Number of Journals&782&\\
   Number of papers &122,514&\\
   Classes& 10&\\
   Research Fields&29&\\\\\
  &\textbf{OMICS}& \textbf{BMC}\\
  \hline
   Number of Journals&498 &334\\
   Number of papers &70,247&369,102\\
   Research Fields &29&18\\
   Total editors&16,859&21,859\\
  Total unique editors&14,665&20,153\\
  Total authors&223,626&2,223,945\\
  Total unique authors&203,143&2,034,394\\\cline{1-3}
 \end{tabular}}
 \caption{Salient statistics about the full OPG, OMICS class in particular, and the BMC Group.}
 \label{tab:dataset}
 \end{minipage}
\end{table}


\section{OMICS publishing group vs. BMC}
\label{sec:analys}
In this section, we analyze the entire publishing eco-system of OPG and compare the interesting findings with BMC publishing group. We leverage several well-known online databases to analyze and authenticate the claims and beliefs.
\subsection{The Meta Information}

\paragraph{\textbf{Indexing in Digital Directories and Scholarly Associations}}
Scholarly digital directories are created with an aim to increase the visibility and ease of use of open access scientific and scholarly journals. Some prominent directories include \textit{SJR} (described in the previous section) and \textit{Directory of Open Access Journals (DOAJ)}. Similarly, several global scholarly associations like \textit{Open Access Scholarly Publishers Association (OASPA)}, \textit{Committee on Publication Ethics (COPE)} and \textit{International Association of Scientific and Technical \& Medical Publishers (STM)}, maintain standards for strict regulations and ethical practices in open access publishing. Only 21 out of 782 ($\sim2.6\%$) journals in OPG are indexed in SJR (see Table~\ref{tab:classes}). Four classes (SciTechnol, RROIJ, Allied Academics and SRL) have no members indexed in SJR. Surprisingly, not a single OPG journal is present in DOAJ. Also, the three associations OASPA, COPE and STM, do not include any OPG journals in their list. 

However, BMC shows a high acceptance in above prominent directories. Out of 334 BMC journals, 288 ($\sim$89\%) journals are indexed in SJR.  Similarly, 322 BMC journals  are  present in DOAJ. Also, the three associations OASPA, COPE and STM, have listed Springer Nature under its members. 

\paragraph{\textbf{Impact Factor}}
Among 782 OPG journals, only 432 journals have reported their impact factor. Out of 432 journals, 69 journals have cited their source agency for impact factor computation. Interestingly, the journal class \textit{SciTechnol} reports impact factor for 39 out of 55 journals and cites \textit{COSMOS}~\cite{cosmos} as a source agency. We found that COSMOS's geographical coordinates points to a kindergarten school in Germany. To our surprise, COSMOS's founder and patron Mr. Jahanas Anderson credentials turned out to be incorrect. A basic Google reverse image search of his profile picture led to a genuine profile of Prof. Stephan Muller\cite{muller_prof}. Interestingly, the certificate of acknowledgment provided by COSMOS contains a barcode that redirects to Food Safety and Standards Regulations of India (FSSAI). In addition, we found that \textit{SciTechnol} falsely claimed to be indexed by reputed services such as Google Scholar, DOAJ, etc. We omit related investigative studies on other classes due to space constraint.

In contrast, BMC leverages services of popular Scopus database~\cite{scopus} for impact factor computation. It also provides several addition citation metrics like SCImago Journal Rank~\cite{scimago}, CiteScore~\cite{citescore}, Altmetric Mentions~\cite{altmet}, etc.

\paragraph{\textbf{Contact Information}}
Next, we studied the availability and authenticity of the postal address of the editorial offices of OPG journals. Interestingly, we find heavy usage of postal mailbox rental service called Prime Secretarial (PS)~\cite{prime}. PS provides facilities for renting private street addresses with an extremely high level of confidentiality. Overall, we found 198 journals that used rented addresses, 500 journals do not provide any postal address information. 48 journals do not have even a contact page. They present a web-based form for official communication. Remaining 36 journals, all from RROIJ class, have provided the postal address of some buildings in the city of Hyderabad, India. 

Similar to OPG, we study the availability and authenticity of the postal addresses of the editorial offices of BMC. None of the BMC journals provide postal address of operating offices. It either provides the email IDs of their employees or redirects to a submission portal.

\paragraph{\textbf{Journal Names}}
Journal names add another dimension to interesting insights. The OPG's journal names are extremely long as opposed to BMC.   
The average number of words and characters in OPG's journal names are 8.3 and 60.2 respectively. 
For BMC, the corresponding values are 3.8 and 29.4. We claim that OPG uses longer names to show authenticity and research coverage. 12 OPG journals from \textit{Trade Science Inc.} subclass are titled as ``ABC: An Indian Journal'' and claim to focus on Indian region where \textit{ABC} includes keywords such as BioChemistry, etc. However, the editorship (10\%) representation from India looks abysmal. In contrast to OPG, we find an entirely different trend in BMC. We find several journals focusing on different countries. All of these journals have high contribution from the their native country. For example, Chinese Neurosurgical Journal, Chinese Medicine, Israel Journal of Health Policy Research, Irish Vetenaray Journal, and Italian Journal of Pediatrics have their respective country contributions as 64\%, 78\%, 70\%, 65\% and  52\% respectively.\\\\
\noindent Next, we perform more nuanced set of experiments with only OMICS subclass (\textit{hereafter as OMICS}) due to the difficulty in data curation (non-uniformity in format, unavailability of paper-specific data, etc.) for several OPG classes. 

\subsection{The Editorial Board}
In this section, we study the editorship information by leveraging editors’ meta-data from the information available at editorial board pages of every journal.

\paragraph{\textbf{Name normalization}}
We find total 16,859 editors in 494 (=34.1 editors per journal)) OMICS journals.\footnote{Four journal editor information links are dead.} As opposed to Beall's hypothesis, the majority of the journals have provided complete information of editors --- name, designation, affiliation and country --- but no contact information such as email, personal homepage, etc. On empirical investigation, we find several instances of editors names with slight variations (acronyms in name, mid name vs. no mid name, etc.) but similar affiliations.\footnote{`Alexander Birbrair' might be present as `A Birbrair' and `Birbrair Alexander'} We normalize these variants by leveraging naming conventions and similarity in affiliations. Normalization results in 14,665 unique editors names. In contrast, we find total 21,859 editors in 334 (=65.4 editors per journal) BMC journals. Similar name normalization scheme, as described above, results in 20,153 unique editors names.

\begin{figure}[t]
\centering
\begin{tabular}{cc}
    \includegraphics[width=0.45\textwidth]{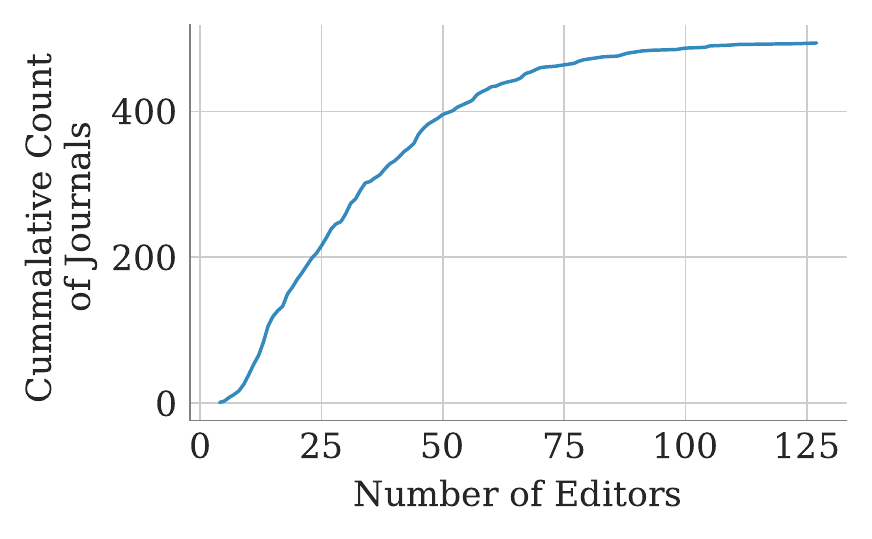}&
    \includegraphics[width=0.45\textwidth]{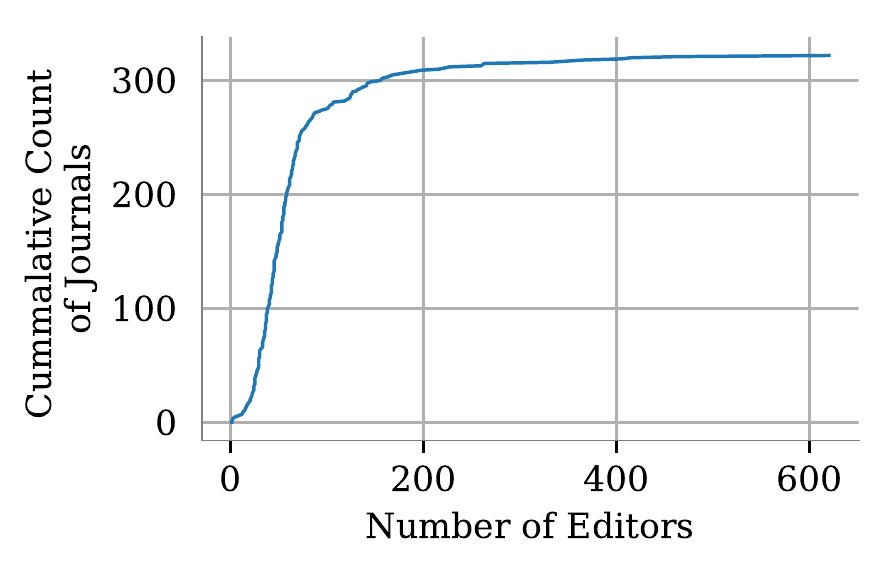}\\
   (a) OMICS & (b) BMC \\
\end{tabular}
\caption{Editor distribution in (a) OMICS subclass, and (b) BMC. 30.9\% OMICS have more than 40 editors. Seven journals have editor count more than 100. (b) 67.06\% BMC journals have more than 40 editors. 45 journals have editor count more than 100.}\label{fig:edit_dist}
\vspace{-0.5cm}
\end{figure}

\paragraph{\textbf{Editor Contribution}} Figure~\ref{fig:edit_dist}a shows editor distribution in OMICS  journals. More than 9\% (=1,409) editors contribute to atleast one journal. 13 editors contribute to more than 10 journals.  30.9\% journals have more than 40 editors. Surprisingly, we find seven journals having editor count more than 100. \textit{`Journal of Tissue Science \& Engineering Open Access Journal`} has maximum editors (=127). On an average, we find 34.1 editors per journal in OMICS. In contrast to the Beall's criterion, the average editors are significantly closer to well-known publishers such as BMC (65.4) and Science (97.8). In BMC (see Figure~\ref{fig:edit_dist}b), more than 6.47\% (=1,316) editors contribute to atleast two journals. No editor contribute to more than 10 journals (maximum count is 7). 67.06\% journals have more than 40 editors. Surprisingly, we find 45 journals having editor count more than 100. \textit{`BMC Public Health`} has maximum number of editors (=620). Figure~\ref{fig:edit_dist}b shows editor distribution in BMC  journals. It is observed that OMICS operates with high journal count and low editorship count while BMC operates with low journals count and high editorship count.
\begin{figure}[!tbh]
\centering
\begin{tabular}{cc}
    \includegraphics[width=0.5\textwidth]{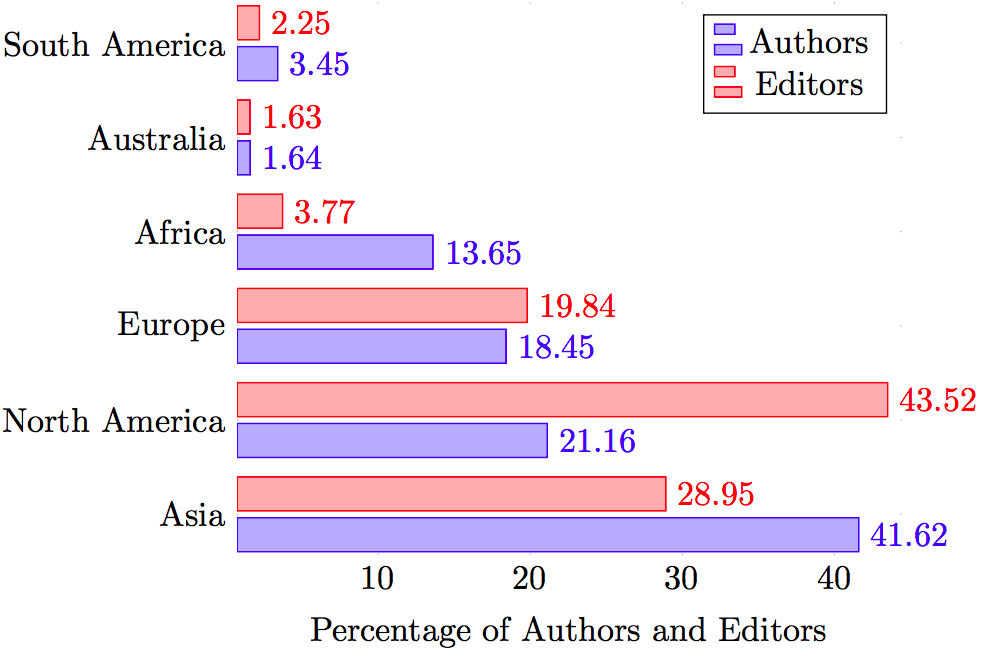}&
    \includegraphics[width=0.5\textwidth]{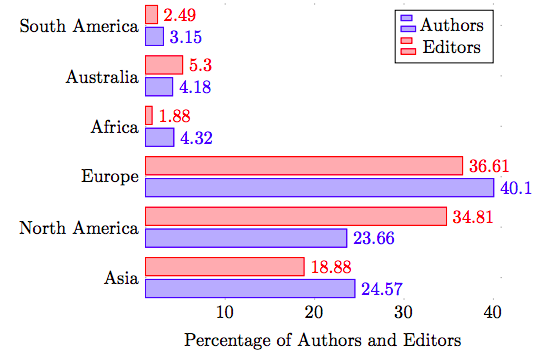}\\
    (a) OMICS - Continent & (b) BMC - Continent\\
    \includegraphics[width=0.5\textwidth]{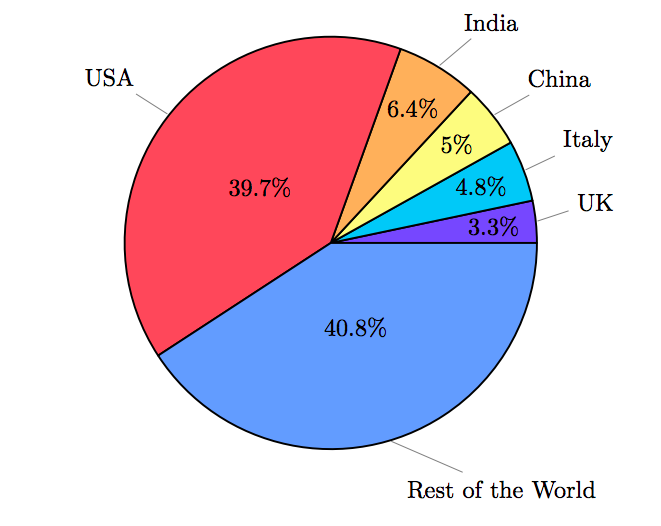}&
    \includegraphics[width=0.45\textwidth]{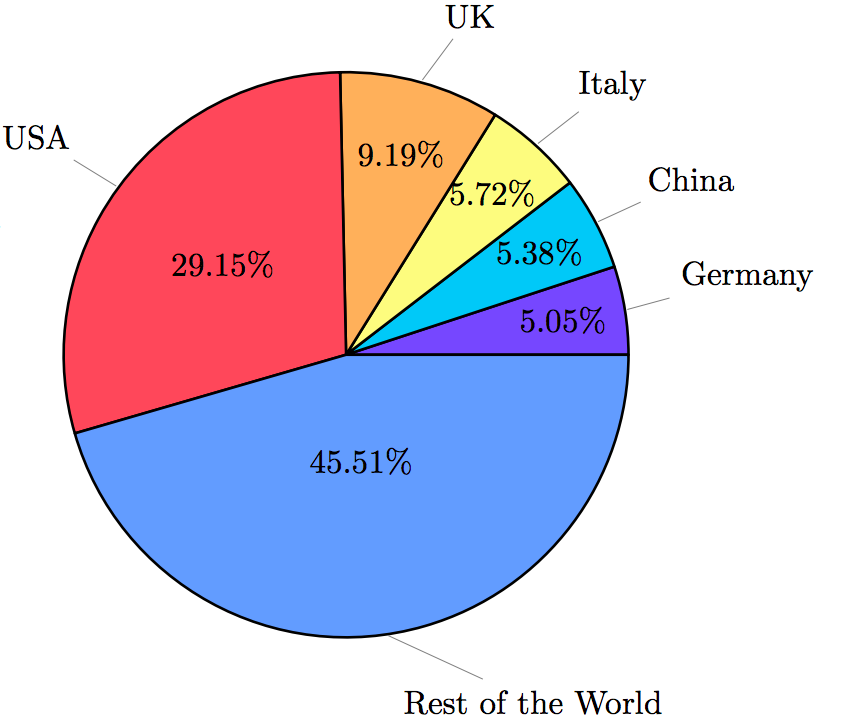}\\
    (c) OMICS - Country & (d) BMC - Country\\
    \includegraphics[width=0.5\textwidth]{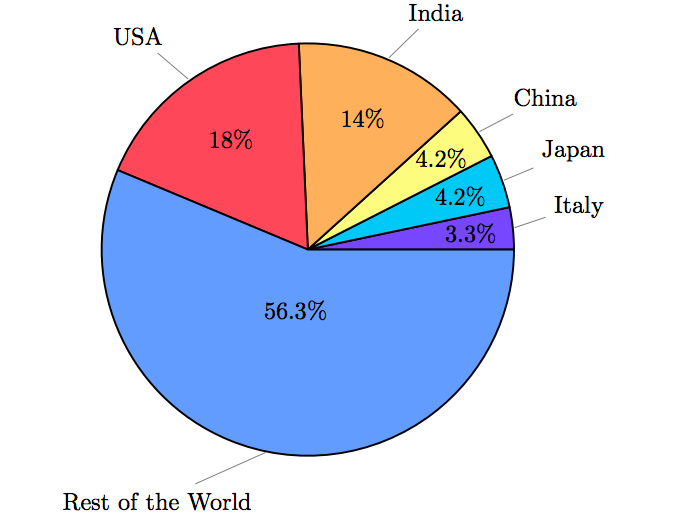}& \includegraphics[width=0.45\textwidth]{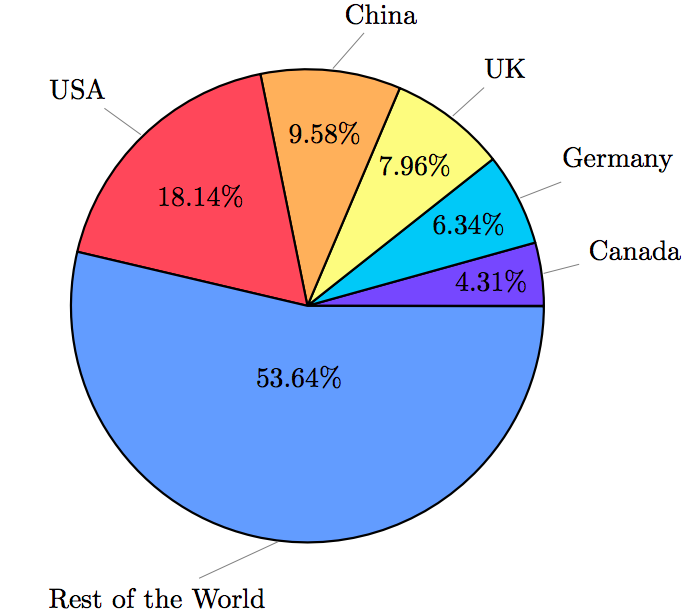}\\
    (e) OMICS - Country & (f) BMC - Country\\
\end{tabular}
\caption{Continent-wise distribution of editors and authors in (a) OMICS and (b) BMC. Country-wise distribution of editors in (c) OMICS and (d) BMC. Country-wise distribution of authors in (e) OMICS and (f) BMC.}\label{fig:geo}
\end{figure}

\paragraph{\textbf{Geographical Analysis}}
We leverage affiliations to analyze the geographical distribution of editors. Figure~\ref{fig:geo}a and ~\ref{fig:geo}c shows geographical distribution of OMICS editors. Continent-wise analysis (see Figure~\ref{fig:geo}a) shows that the majority of editors are affiliated to North American organizations (43.52\%) followed by Asia (28.95\%). Similar country-wise analysis show that editor affiliates to more than 131 countries across the world (see Figure~\ref{fig:geo}b), with USA being the major contributor (39.7\%) followed by India (6.4\%), China (5\%), Italy (4.8\%) and UK (3.3\%). In BMC, continent-wise analysis (see Figure~\ref{fig:geo}b) shows that the majority of authors affiliates to European organizations (36.61\%) followed by North America (34.81\%). Editors belong to more than 136 countries across the world (see Figure~\ref{fig:geo}d), with USA being the major contributor (29.15\%) followed by UK (9.19\%), Italy (5.72\%), China (5.38\%) and Germany (5.05\%). In contrast to OMICS, Indian editors have marginal contribution (1.51\%). It is observed that both of the groups show similar geographical diversity in editor contribution.

\paragraph{\textbf{Gender Analysis}}
Next, we leverage popular gender prediction tool \textit{Gender Guesser}~\cite{gender_guess} to infer gender bias information from the names of the editors in OMICS and BMC. Given a name, the tools outputs a probability of its gender. The probability of an editor being a male came out to be 0.76. We find that 467 journals were heavily male dominated (P(male)$>$0.5). 36 journals among 467, do not include any female member in the editorial board. In case of BMC, The probability of an editor being a male came out to be 0.72. We find that 302 journals were heavily male dominated (P(male)$>$0.5). 12 journals among 302, do not include any female member in the editorial board. 

\subsection{The Authorship}
In this section, we study the authorship information by leveraging authors' metadata from paper headers in the journal's archive. 
\paragraph{\textbf{Name Normalization}}
In majority of the papers published in OMICS, we obtained well-structured metadata information--- affiliation with the department/lab/center, the organization, city, and country name. As opposed to editors, phone number and email information of the first author is also available. We leverage this metadata information to normalize and index authors (similar to the editor name normalization). In OMICS, normalization results in 203,143 unique authors out of 223,626 authors. Similarly, BMC comprises 2,034,394 unique authors among 2,223,945 authors. 

\begin{figure}[t]
\centering
\begin{tabular}{cc}
     \includegraphics[width=0.45\textwidth]{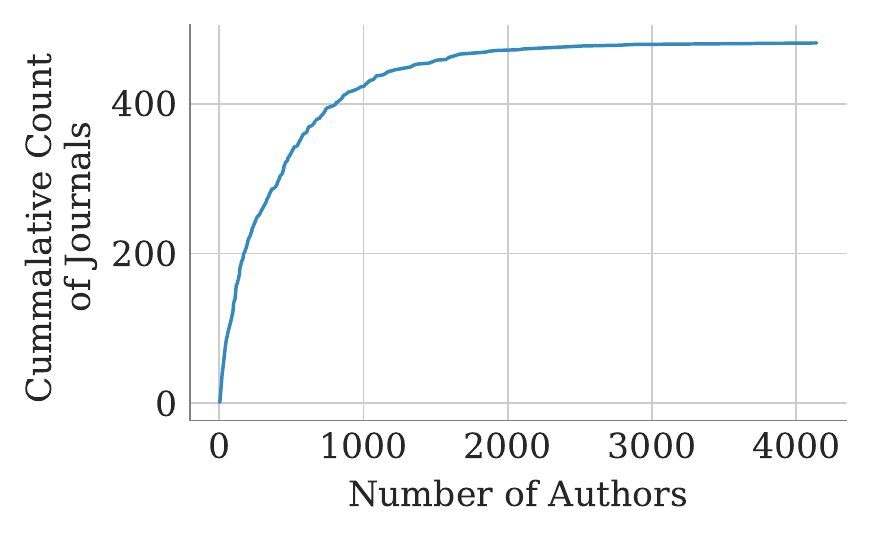}&
     \includegraphics[width=0.45\textwidth]{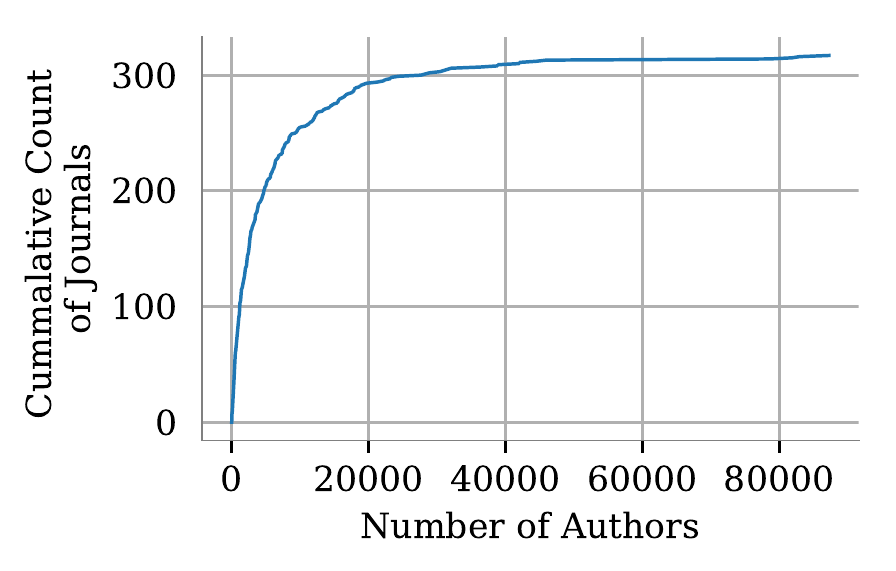}\\
   (a) OMICS & (b) BMC \\
\end{tabular}
\caption{Authorship distribution in (a) OMICS, and (b) BMC. 6.7\% OMICS journals have published in more than one journal. 6.3\% authors in BMC journals published in more than one journal.}\label{fig:auth_dist}
\vspace{-0.5cm}
\end{figure}

\paragraph{\textbf{Author Contribution}}
Among 498 OMICS journals, the author information was available in 481 ($\sim$96\%) journals.  The rest of the journals do not possess an archive portal (12 cases), points to the same archive link (three cases), or contains only volume names but no paper information (two case). Figure~\ref{fig:auth_dist}a shows author publishing statistics. 6.7\% authors published in more than one journal. We found 20 authors that published in more than 10 journals. 97 authors published in more than five journals. Similar to OMICS, BMC has the author information availability in 318 ($\sim$95\%) journals. The remaining journals (=16) have not started published digitally yet. Figure~\ref{fig:auth_dist}b shows author publishing statistics for BMC. 6.3\% authors published in more than one journal. We found 558 authors that published in more than 10 different journals. 

\paragraph{\textbf{Geographic Analysis}}
Next, we perform a geographical analysis of the authors. In contrast with editors, here we study the geographical distribution of the first author of the paper by leveraging the affiliation information. Contrary to the OMICS CEO's claims~\cite{yadav_2018}, 
that \textit{'Our articles from India are very few, less than 0.01 \% -- 99.99 \% of articles are from outside'}, we find that Indian authorship is $\sim$14\% of the global authorship. Also, we find that majority of authors publishing in OMICS class are from Asia(41.62\%) (see Figure~\ref{fig:geo}a and~\ref{fig:geo}e). In BMC, majority of the authorship comes from Europe(40.1\%) followed by Asia(23.66\%) and North America(24.57\%). It is observed that the combined authorship from three continents, Asia, Africa and South America, having higher percentage of developing countries~\cite{un}  contributes more towards OMICS than BMC.

\paragraph{\textbf{Gender Analysis}}
We perform gender analysis of authors by inferring gender from author names (similar to editor gender identification). In OMICS, the probability of an author being a male came out to be 0.64. We found that 432 OMICS journals were heavily male dominated (P(male) $>$ 0.5) and eight among 432 journals have no contribution from any female author. Surprisingly, in case of BMC, the probability of an author being a male came out to be 0.64. We found that 283 journals were heavily male dominated. However, each journal has non-zero female author count.

\subsection{Novel data-driven signals}
In contrast to previously proposed subjective list-based evaluation techniques, we can leverage various co-authorship signals and network-centric properties to identify distinctive features between predatory and reputed publishers. 

\paragraph{\textbf{Author-Editor Commonness}}
OMICS shows significant fraction of editors that published their own work in their journal. 3.2\% authors are the editors in the same journal. This phenomenon was observed across 426 Journals out of 498 Journal with an average of 16.8 editors per journal. 4.6\% editors are authors in atleast one of the OMICS class journal other than the journal they are editor in. In total, an average of 37.3 editors publish in each journal of OMICS group. In contrast, above trend is negligible in BMC. 0.2\% authors are the editors in the same journal. 

\paragraph{\textbf{Editor Network}}
Next, we perform network analysis over the editor graph. Editor graph $G_E(V,E)$ comprises editors as nodes $V$, whereas edges $E$ connect editors in the same journal. We rank nodes based on three centrality measures: (i) Degree, (ii) Eigenvector, and (iii) Betweenness.  The degree centrality ranks nodes with more connections higher in terms of centrality. A more popular editor has a high degree. However, in real-world scenarios, having more co-editors does not by itself guarantee that some editor is popular. We, therefore, also compute Eigenvector centrality which measures editor connections with other popular (not just any other editor) editors. Betweenness centrality measures how often a node occurs on all shortest paths between two nodes. In the current graph, it measures how frequently editorial decisions/recommendations are passed through that node. Overall, $G_E$ of OMICS contains 14,665 nodes and 401,063 edges.  The profiles of top-ranked editors (based on three metrics are verified by looking for their Google Scholar and university/organization webpage. Interestingly, the fourth-ranked editor as per degree centrality  named \textit{'Alireza Heidari'} belongs to 'California South University'~\cite{csu_home} which is claimed to be fake by Huffington Post~\cite{shakeri_2018}. 
Also, an editor named \textit{'Sasikanth Adigopula'} in \textit{Journal of Cardiovascular Diseases and Diagnosis}~\cite{jour_omics} is claimed to be affiliated from Banner University, Colorado. On researching through internet sources, we found that there is no such institution and the name is of a cardiologist from Loma Linda University. 

\begin{table}[!tbh]
\resizebox{\hsize}{!}{
\begin{tabular}{|c|c|c||c|c|c|}
\hline
\multicolumn{3}{|c|}{OMICS}&\multicolumn{3}{|c|}{BMC}\\\hline
\bf Editor Name & \bf  Centrality Values & \bf Journal Count& \bf Editor Name & \bf  Centrality Values & \bf Journal Count\\\hline
\multicolumn{6}{|c|}{Degree}\\\hline
George Perry  & 0.033 & 26&Sang Yup Lee  & 0.033 &  7 \\
Rabiul Ahasan &  0.059  & 20 &Stefano Petti  & 0.028 &  2 \\
Kenneth Maiese &0.051&10 &Guy Brock  & 0.025 &  4 \\
Alireza Heidari &0.045&13 &Lalit Dandona  & 0.024 &  4 \\
Rajesh Rajan Wakaskar  & 0.044  & 11 &John Ioannidis  & 0.023 &  5 \\\hline
\multicolumn{6}{|c|}{Eigenvector}\\\hline
George Perry & 0.001 & 26 &Guy Brock  & 0.054 &  4 \\
Rabiul Ahasan & 0.0009 & 20 &Stefano Petti  & 0.053 &  2 \\
Sandeep Kumar Kar & 0.0006 & 14 &Jonathan Mant  & 0.053 &  3 \\
Alexander Birbrair & 0.0006 & 14 &Shazia Jamshed  & 0.051 &  4 \\
Alireza Heidari & 0.0006 & 13 &Florian Fischer  & 0.049 &  3 \\\hline
\multicolumn{6}{|c|}{Betweenness}\\\hline
George Perry & 0.100 & 26 &Sang Yup Lee  & 0.0002 &  7 \\
Rabiul Ahasan & 0.088 & 20 &Fernando Schmitt  & 0.0002 &  5 \\
Sandeep Kumar Kar & 0.039 & 14 &Jean-Louis Vincent & 0.0002 &  5 \\
Akbar Nikkhah & 0.030 & 11 &Suowen Xu  & 0.0001&  5 \\
Alexander Birbrair & 0.030 & 14 &Yong Wang  & 0.0001&  5 \\\hline
\end{tabular}}
\caption{Top-5 central nodes in editor graph based on three centrality measure.}\label{tab:editor}
\vspace{-0.5cm}
\end{table}

Similar graph construction for BMC editors yields 20,153 nodes and 1,419,776 edges. However, BMC shows no dubious signals in the top-ranked nodes. The editors either possess GS or ORCID user profiles or have authentic university profile pages. Detailed centrality values for Top-5 editors in each measure is shown in Table \ref{tab:editor}.

\section{The Evolving Ecosystem}
\label{sec:evol}
The previous section compares potential predatory publishing group OMICS with popular publishing group BMC. In this section, we revisit Beall's list of subjective criteria for predatory classification. Table~\ref{tab:compare_beall} shows a comparison between OMICS and BMC against the list of 35 Beall's criteria that can be easily verified through internet resources with a minimum requirement for manual processing. 
Some of the interesting insights are:
\begin{itemize}[noitemsep,nosep]
    \item 22 criteria are common between OMICS and BMC. 
    \begin{itemize}[noitemsep,nosep]
        \item Five criteria are satisfied by both OMICS and  BMC.
    \end{itemize}
    \item 13 criteria are satisfied by OMICS but not by BMC.
    \item No criteria being satisfied by BMC but not by OMICS.\\
\end{itemize}

The predatory ecosystem is cautiously changing. The evolution in OMICS results in its operations similar to BMC publishing group. For example, recently, OMICS started its online submission portal similar to well-known publishers (Nature, BMC, and Science). Earlier, it accepts manuscripts through email. It is becoming extensively hard to distinguish between authentic and predatory journals using a standard list of criteria or rules. We need more data-driven identification methods that can detect false/misleading information in publishing groups. Some of the important freely available resources are Google Scholar API\cite{scholar}, Scopus API\cite{scop}, image search facility\cite{google}and DOI/ISSN API\cite{cross}\cite{issn}.

\begin{table}[t]
\begin{center}
\resizebox{\hsize}{!}{
\begin{tabular}{|c|c|c|}
\hline
\bf Beall's Criteria & OMICS & BMC\\
\hline
 \parbox[t]{13cm}{The publisher's owner is identified as the editor of each and every
journal published by the organization.} & \cellcolor{red!25}$\times$ &\cellcolor{red!25}$\times$ \\ \hline
 \parbox[t]{13cm}{No single individual is identified as any specific journal's editor.}& \cellcolor{red!25}$\times$ &\cellcolor{red!25}$\times$  \\ \hline
 \parbox[t]{13cm}{The journal does not identify a formal editorial / review board.}& \cellcolor{red!25}$\times$ &\cellcolor{red!25}$\times$  \\ \hline
 \parbox[t]{13cm}{No academic information is provided regarding the editor, editorial staff, and/or review board members (e.g., institutional affiliation).} & \cellcolor{red!25}$\times$ &\cellcolor{red!25}$\times$ \\ \hline
 \parbox[t]{13cm}{Evidence exists showing that the editor and/or review board members do not possess academic expertise in the journal's field.}& \cellcolor{green!25}$\checkmark$ &\cellcolor{red!25}$\times$  \\ \hline
\parbox[t]{13cm}{Two or more journals have duplicate editorial boards} & \cellcolor{red!25}$\times$ &\cellcolor{red!25}$\times$\\ \hline
 \parbox[t]{13cm}{The journals have an insufficient number of board members} & \cellcolor{red!25}$\times$ &\cellcolor{red!25}$\times$  \\ \hline
 \parbox[t]{13cm}{The journals have concocted editorial boards (made up names)}& \cellcolor{green!25}$\checkmark$ &\cellcolor{red!25}$\times$  \\ \hline
 \parbox[t]{13cm}{The editorial board engages in gender bias (i.e., exclusion of any female members)} & \cellcolor{green!25}$\checkmark$ &\cellcolor{green!25}$\checkmark$ \\ \hline
 \parbox[t]{13cm}{Has no policies or practices for digital preservation, meaning that if the journal ceases operations, all of the content disappears from the internet.} & \cellcolor{green!25}$\checkmark$ &\cellcolor{red!25}$\times$ \\ \hline
 \parbox[t]{13cm}{Begins operations with a large fleet of journals, often using a common template to quickly create each journal's home page.} & \cellcolor{green!25}$\checkmark$ &\cellcolor{green!25}$\checkmark$\\ \hline
 \parbox[t]{13cm}{Does not allow search engines to crawl the published content, preventing the content from being indexed in academic indexes.}& \cellcolor{red!25}$\times$ &\cellcolor{red!25}$\times$  \\ \hline
 \parbox[t]{13cm}{Copy-proofs (locks) their PDFs, thus making it harder to check for plagiarism.}& \cellcolor{red!25}$\times$ &\cellcolor{red!25}$\times$  \\ \hline
 \parbox[t]{13cm}{On its website, the publisher falsely claims one or more of its journals have actual impact factors, or advertises impact factors assigned by fake ``impact factor'' services, or
it uses some made up measure} & \cellcolor{green!25}$\checkmark$ &\cellcolor{red!25}$\times$ \\ \hline
 \parbox[t]{13cm}{The publisher falsely claims to have its content indexed in legitimate abstracting and indexing services or claims that its content is indexed in resources that are not abstracting and indexing services.} & \cellcolor{green!25}$\checkmark$ &\cellcolor{red!25}$\times$ \\ \hline
 \parbox[t]{13cm}{Use boastful language claiming to be a ``leading publisher'' even though the publisher may only be a startup or a novice organization.}& \cellcolor{green!25}$\checkmark$ &\cellcolor{green!25}$\checkmark$ \\ \hline
 \parbox[t]{13cm}{Operate in a Western country chiefly for the purpose of functioning as a vanity press for scholars in a developing country (e.g., utilizing a maildrop address or PO box address in the United States, while actually operating from a developing country).}& \cellcolor{green!25}$\checkmark$ &\cellcolor{red!25}$\times$  \\ \hline
 \parbox[t]{13cm}{Have a ``contact us'' page that only includes a web form or an email address, and the publisher hides or does not reveal its location.}& \cellcolor{green!25}$\checkmark$ &\cellcolor{green!25}$\checkmark$  \\ \hline
 \parbox[t]{13cm}{The publisher lists insufficient contact information, including contact information that does not clearly state the headquarters location or misrepresents the headquarters location (e.g., through the use of addresses that are actually mail drops).}& \cellcolor{green!25}$\checkmark$ &\cellcolor{green!25}$\checkmark$  \\ \hline
 \parbox[t]{13cm}{The publisher publishes journals that are excessively broad (e.g., Journal of Education) in order to attract more articles and gain more revenue from author fees.}& \cellcolor{green!25}$\checkmark$ &\cellcolor{red!25}$\times$  \\ \hline
 \parbox[t]{13cm}{The publisher publishes journals that combine two or more fields not normally treated together (e.g., International Journal of Business, Humanities and Technology)} & \cellcolor{red!25}$\times$ &\cellcolor{red!25}$\times$ \\ \hline
  \parbox[t]{13cm}{The name of a journal does not adequately reflect its origin (e.g., a journal with the word ``Canadian'' in its name when neither the publisher, editor, nor any purported institutional affiliate relates whatsoever to Canada).} & \cellcolor{red!25}$\times$ &\cellcolor{red!25}$\times$ \\ \hline
 \parbox[t]{13cm}{The publisher has poorly maintained websites, including dead links, prominent misspellings and grammatical errors on the website.} & \cellcolor{green!25}$\checkmark$ &\cellcolor{red!25}$\times$ \\ \hline
 \parbox[t]{13cm}{The publisher makes unauthorized use of licensed images on their website, taken from the open web, without permission or licensing from the copyright owners.}& \cellcolor{green!25}$\checkmark$ &\cellcolor{red!25}$\times$ \\ \hline
 \parbox[t]{13cm}{The publisher does not use standard identifiers (ISSN/DOI) or uses them improperly.} & \cellcolor{green!25}$\checkmark$ &\cellcolor{red!25}$\times$ \\ \hline
 \parbox[t]{13cm}{The publisher uses names such as ``Network,'' ``Center,'' ``Association,'' ``Institute,'' and the like when it is only a solitary, proprietary operation and does not meet the definition of the term used or implied non-profit mission.} & \cellcolor{red!25}$\times$ &\cellcolor{red!25}$\times$ \\ \hline
 \parbox[t]{13cm}{The publisher has excessive, cluttered advertising on its site to the extent that it interferes with site navigation and content access.}& \cellcolor{red!25}$\times$ &\cellcolor{red!25}$\times$  \\ \hline
 \parbox[t]{13cm}{The publisher has no membership in industry associations and/or intentionally fails to follow industry standards.} & \cellcolor{green!25}$\checkmark$ &\cellcolor{red!25}$\times$ \\ \hline
 \parbox[t]{13cm}{The publisher includes links to legitimate conferences and associations on its main website, as if to borrow from other organizations’ legitimacy, and emblazon the new publisher with the others' legacy value.}& \cellcolor{red!25}$\times$ &\cellcolor{red!25}$\times$  \\ \hline
 \parbox[t]{13cm}{The publisher or its journals are not listed in standard periodical directories or are not widely cataloged in library databases.}& \cellcolor{green!25}$\checkmark$ &\cellcolor{red!25}$\times$  \\ \hline
 \parbox[t]{13cm}{None of the members of a particular journal's editorial board have ever published an article in the journal.} & \cellcolor{red!25}$\times$ &\cellcolor{red!25}$\times$ \\ \hline
 \parbox[t]{13cm}{There is little or no geographic diversity among the authors of articles in one or more of the publisher's journals, an indication the journal has become an easy outlet for authors from one country or region to get scholarly publications.} & \cellcolor{red!25}$\times$ &\cellcolor{red!25}$\times$ \\ \hline
 \parbox[t]{13cm}{The publishers' officers use email addresses that end in .gmail.com, yahoo.com, or some other free email supplier.} & \cellcolor{red!25}$\times$ &\cellcolor{red!25}$\times$ \\ \hline
 \parbox[t]{13cm}{The publisher displays prominent statements that promise rapid publication and/or unusually quick peer review.} & \cellcolor{red!25}$\times$ &\cellcolor{red!25}$\times$ \\ \hline
 \parbox[t]{13cm}{The publisher copies ``authors guidelines'' verbatim from other publishers.}& \cellcolor{green!25}$\checkmark$ &\cellcolor{red!25}$\times$\\ \hline
\end{tabular}}
\end{center}
\caption{Comparison between OMICS and BMC against Beall's subjective criterion. }\label{tab:compare_beall}
\end{table}

\section{Conclusion and Future Work}
\label{sec:end}
In this work, we present an empirical analysis of a potential predatory open access publisher OMICS and compare it with well-known and highly reputed publisher BMC. We present facts with substantial evidence gathered from reputable sources to refute popular claims. We show that the entire predatory ecosystem is cautiously evolving to bypass the standard filters. Because the current work is only a preliminary attempt to study India-based potential predatory publishers extensively, future extensions could lead to similar studies on all possible dubious journals along with the development of web interfaces to visually represent these evidence. 

\clearpage
\bibliography{ref}
\end{document}